\documentclass[pdflatex,sn-mathphys-num]{sn-jnl}

\usepackage{graphicx}%
\usepackage{multirow}%
\usepackage{amsmath,amssymb,amsfonts}%
\usepackage{amsthm}%
\usepackage{mathrsfs}%
\usepackage[title]{appendix}%
\usepackage{xcolor}%
\usepackage{textcomp}%
\usepackage{manyfoot}%
\usepackage{booktabs}%
\usepackage{algorithm}%
\usepackage{algorithmicx}%
\usepackage{algpseudocode}%
\usepackage{listings}%
\usepackage{longtable}

\usepackage{pdflscape}

\theoremstyle{thmstyleone}%
\theoremstyle{thmstyletwo}%

\theoremstyle{thmstylethree}%

\begin{document}

\title[Magnetic Moment-Field Interactions] {Magnetic Moment-Field Interactions: A Universal Mechanism for Particle Energization}


\author*[1]{\fnm{Anil } \sur{Raghav}}\email{anil.raghav@physics.mu.ac.in}

\author[1]{\fnm{Ajay} \sur{Kumar}}

\author[1]{\fnm{Mariyam} \sur{Karari}}
\author[1]{\fnm{Shubham} \sur{Kadam}}
\author[1]{\fnm{Kalpesh} \sur{Ghag}}
\author[1]{\fnm{Kishor} \sur{Kumbhar}}
\author[1]{\fnm{Omkar} \sur{Dhamane}}
\affil[1]{\orgdiv{Department of Physics}, \orgname{University of Mumbai}, \orgaddress{\street{Santacruz}, \city{Mumbai}, \postcode{400098}, \state{Maharashtra}, \country{India}}}



\abstract{
Magnetic reconnection is a pivotal mechanisms in the energization and heating of cosmic plasmas, yet the exact process of energy transfer during these events remain elusive. Traditional models, which focus on acoustic and magnetohydrodynamic waves and micro/nano-flares, fall short of explaining the extreme heating of the solar corona and the origins of the supersonic solar wind. In this study, we provide compelling observational evidence from Wind spacecraft data supporting the Raghav effect—a mechanism where interactions between the magnetic moments of charged particles and dynamic magnetic fields result in abrupt kinetic energy changes. Our analysis demonstrates that the observed proton plasma heating is consistent with theoretical predictions, establishing the Raghav effect as a universal mechanism for particle energization. This discovery offers a unified framework for understanding energy dynamics across a wide range of astrophysical magnetised plasma environments.}




\maketitle

\section{Introduction} \label{sec:intro}

The study of plasma heating mechanisms is pivotal in astrophysics, offering insights into energy dissipation in cosmic environments. A key challenge is understanding the anomalously high temperatures in certain astronomical plasmas, such as the solar corona, compared to the cooler underlying photosphere and chromosphere \cite{klimchuk2006solving, aschwanden2006physics}.
This temperature discrepancy, known as the coronal heating problem, has driven extensive theoretical and observational efforts to elucidate the underlying physics. 
In addition, the solar wind presents another puzzle, undergoing continuous heating and acceleration as it emanates from the Sun, which cannot be fully explained by classical thermal expansion models \cite{mccomas2007understanding,cranmer2007self,cranmer2019properties}. This suggests the involvement of non-adiabatic processes that efficiently transfer energy across various scales within the plasma.

Plasma heating mechanisms are diverse, encompassing both collisional and collisionless processes. Ohmic (resistive) heating, driven by finite electrical resistivity, and magnetic reconnection events, such as Parker's `nanoflares,' are fundamental contributors to energy dissipation in magnetized plasmas \cite{lin1984solar,parker1988nanoflares,shibata2011solar, spitzer2006physics}. Additionally, magnetohydrodynamic (MHD) waves, particularly Alfvén waves, transport energy from the dense lower atmosphere into the corona, where their dissipation through resonant absorption, phase mixing, and turbulent cascades contributes to heating \cite{osterbrock1961heating,heyvaerts1983coronal,davila1987heating,sakurai1984generation}. 
At sub-ion scales, kinetic processes and wave-particle interactions significantly affect the temperature distribution of electrons and ions \cite{gary1993theory,grovselj2019kinetic}. The continuous increase in temperature and acceleration of the solar wind beyond simple predictions of the conductive model are attributed to turbulent dissipation, wave-particle interactions, and kinetic instabilities in the expanding plasma \cite{marsch2006kinetic}. Recent studies indicate that cascading of MHD turbulence to kinetic scales provides a persistent source of heating necessary to maintain the observed solar wind properties \cite{bruno2013solar}. These multi-scale processes collectively form a complex picture of plasma heating, challenging our understanding of energy transfer in space and astrophysical plasmas.

The persistent gap between theoretical predictions and observational data highlights the need for new plasma-heating mechanisms. Traditional models, focusing on waves, turbulence, and magnetic reconnection, have not fully captured the complexity of observed heating phenomena. Raghav (2025) introduces an innovative mechanism in which interactions between charged-particle magnetic moments and magnetic-field reconfigurations drive plasma heating \cite{raghav2025solarenigmadecodingmystery}. This approach directly links magnetic topology with energy transfer at the particle scale, potentially providing a unified explanation for both coronal heating and solar-wind acceleration. Although the Raghav effect shows promise, it requires further experimental and observational validation to establish its role as a universal heating process in magnetized plasmas. This study demonstrates observational evidence that strongly supports Raghav's hypothesis, marking a significant step forward in understanding plasma heating dynamics. 

\section{Raghav Effect}
 In classical theory, the magnetic moment ($\mu$) is considered an adiabatic invariant in a slowly varying magnetic field, and it is defined as:

\begin{equation}
\mu = \frac{\frac{1}{2}m v_\perp^2}{B},
\end{equation}

where, $v_\perp$ is the velocity perpendicular to the magnetic field ($B$) and $m$ is the mass of the charged particle. The kinetic energy and thermal energy are related through the kinetic theory of gases, expressed as:

\begin{equation}
\frac{1}{2}m v_\perp^2 = \frac{1}{2} K T,
\end{equation}
where $T$ is the particle temperature, and $K$ is the Boltzmann constant. Rearranging these terms yields:
\begin{equation}
 T = 2 \frac{\mu}{K} B
\end{equation}

Before a reconnection event, $\mu$ acts as a stable adiabatic invariant, providing a well-defined initial condition for the system. During reconnection, the magnetic field configuration changes rapidly, yet particles retain the memory of their previously conserved magnetic moment. As the field reconfigures, it exerts a torque on the particles, aligning their magnetic moments antiparallel to the new orientation of field lines. This realignment results in work being done on the particles, leading to an energy increase. If the magnetic field changes by $\delta B$, the corresponding temperature gain is given by,

\begin{equation}
 \delta T = 2 \frac{\mu}{K} \delta B
\end{equation}

Dividing this by the expression for T, we find:
\begin{equation}
\frac{\delta T}{T}  = \frac{\delta B}{ B}
\end{equation}

Thus, the Raghav effect suggests an imperative relationship between temperature gain and magnetic field changes. It implies that the relative (percentage) gain in temperature is directly equivalent to the relative (percentage) decrease in the magnetic field. A key challenge in our study was the assumption that the magnetic moment remains constant before and during the reconnection process. This assumption is not valid by the fact that, after reconnection, particles gain energy. To test this hypothesis, we analyzed in-situ data from events where the magnetic field decreases. These events are critical for understanding the relationship between magnetic field changes and particle energization. By closely monitoring temperature increases during these transitions, we aimed to verify whether the observed energy transfer aligns with our hypothesis regarding the behavior of the $\mu$ during reconnection.

\section{Observation}

To validate the proposed mechanism, it is essential to conduct an initial demonstration. Direct measurements of the solar wind provide a concrete method for testing this hypothesis. For our study, we chose data from June 2008 to May 2009, a period characterized by solar minimum conditions, to reduce the impact of solar transients on the ambient solar wind.

We utilized high-resolution data, recorded every three seconds, from the Magnetic Field Investigation (MFI) \cite{lepping1995wind} and the 3D Plasma and Energetic Particle (3DP) \cite{lin1995three} instruments aboard the Wind spacecraft to examine the solar wind environment. Our study focused on events characterized by a distinct increase in plasma temperature concurrent with a decrease in magnetic field strength. To ensure the integrity of our analysis, we deliberately excluded events marked by abrupt changes, such as shocks and discontinuities. This careful selection process led to the identification of 168 pertinent events. We examined the associated plasma parameters in detail to better understand the interplanetary conditions during these events. We further quantified the observed changes by calculating the percentage decrease in total magnetic field strength and the corresponding percentage increase in plasma temperature. The algorithm and quantification method for this estimation are described in the appendix.

\subsection{Example event}
We illustrate a representative event featuring two potential reconnection site transits, as depicted in Figure 1. The first crossing, marked in cyan, occurs on June 4, 2008, from 16:26:00 UT to 16:28:06 UT, while the second crossing, highlighted in orange, takes place from 16:28:42 UT to 16:29:42 UT. The top panel displays the temporal variation of the total interplanetary magnetic field (IMF) strength (\textbf{B}). During the first interval, we observed a sharp decrease in \textbf{B} from approximately 2.59 nT to a minimum of 1.32 nT, with all IMF components (Bx, By, Bz) showing corresponding variations. The percentage change in the magnetic field is 48.87\%. Interestingly, the solar wind speed decreased by 10 km/s. In contrast, there was a simultaneous increase in temperature, plasma proton density, and plasma beta, indicating dynamic changes in the plasma environment. Specifically, the plasma temperature rose sharply from a minimum of 59,166 K to 88,022 K, resulting in a percentage increase in temperature of about 48.77\%. In the second interval, \textbf{B} decreases from about 2.58 nT to a minimum of 1.82 nT, a 29.38\% change, with a 7 km/s decrease in solar wind speed. Again, there is an increase in temperature, plasma proton density, and plasma beta. The plasma temperature rises from 61,756 K to 79,319 K, marking a 28.44\% increase.

 \begin{figure*}
 \includegraphics[width=1\textwidth]{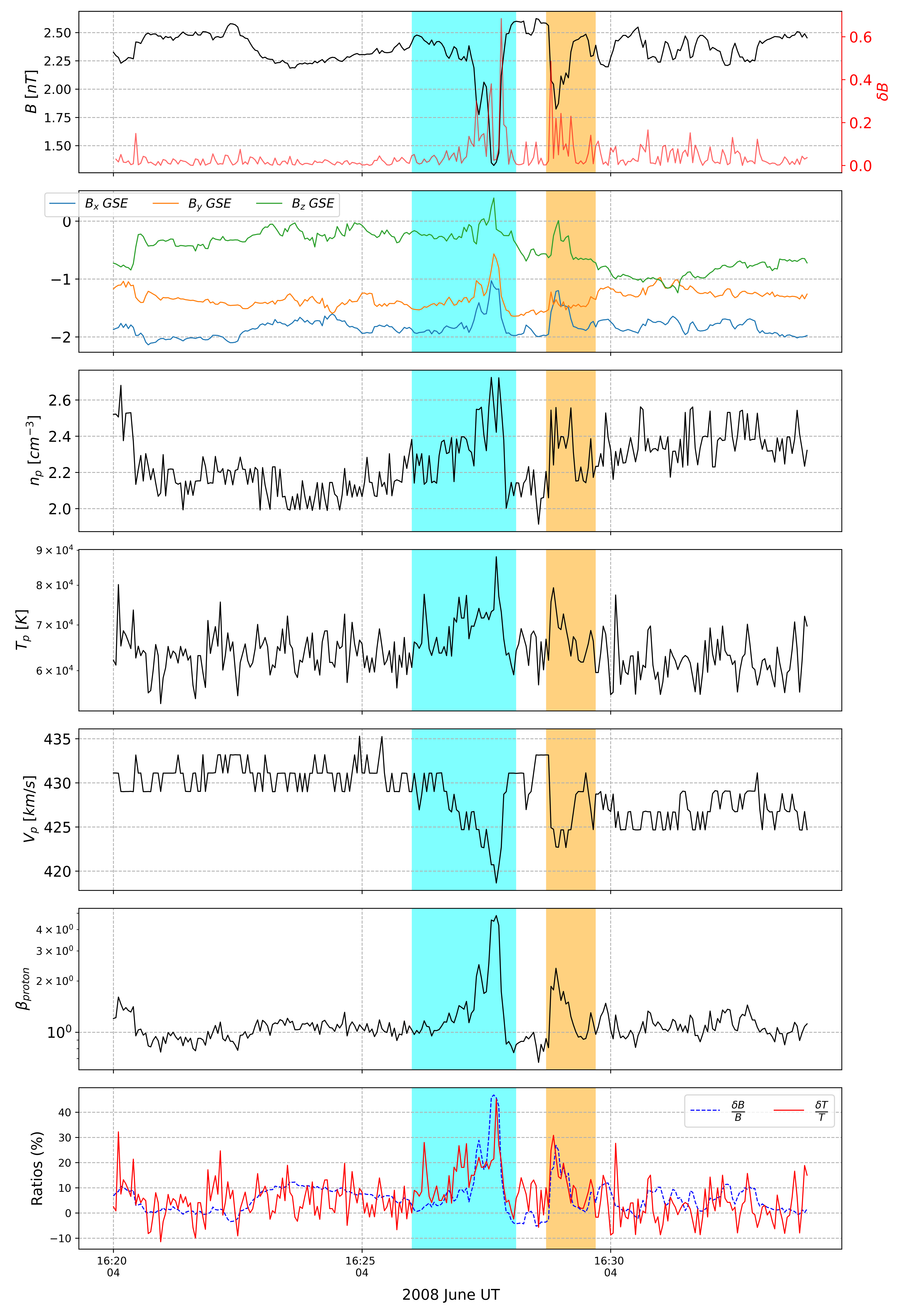}
  \caption{Interplanetary parameters for the prototype event observed by the Wind spacecraft on June 04, 2008, from 16:20 UT to 16:34 UT. From top to bottom, the panels display the magnetic field strength (\textbf{B}) and its fluctuations, magnetic field components, proton number density, proton temperature, solar wind proton velocity, and plasma beta. The bottom panel illustrates the relative change in the magnetic field (inverted mode) and temperature. The cyan and orange shaded regions highlight the area of interest. } 
 \end{figure*}
 
We found remarkable coherence between the magnetic field and plasma temperature variation during the event. Specifically, the percentage decrease in the magnetic field strength ($48.87 \%~\&~ 29.38 \%$) closely matches the increase in plasma temperature ($48.77 \% ~\&~ 28.44\%$), supporting our initial hypothesis. However, relying on a single event can sometimes lead to conclusions based on chance or coincidence. To ensure the robustness of our results, we extended our analysis to multiple events, aiming for statistical significance.

The accompanying document provides a detailed list of 163 events, along with the respective changes in magnetic field and temperature. Figure 2 illustrates the relationship between percentage variations for both parameters. We observed variations in magnetic field and temperature ranging from 20 \% to 94\%. Although these magnetic decrease events occur near possible reconnection sites, the observed variations depend significantly on the spacecraft's distance from the null point during crossover. Each blue dot represents a data point, and the red line is the least square fit, showing a nearly perfect linear correlation with a slope of 0.98. The high correlation coefficient ($R^2=0.99$) suggests a strong linear relationship, indicating that temperature changes are closely associated with changes in the magnetic field. This relationship supports the Raghav hypothesis that magnetic field variations significantly impact temperature dynamics in the studied environment.

In addition, we identified events involving thick current sheets, heliospheric current sheets, boundary sectors, and sharp discontinuities. In these scenarios, the temperature increase is much more significant than the decrease or increase in magnetic field strength. We believe these structures formed well before the spacecraft's transit, allowing particles to cross these structures multiple times, thereby significantly increasing the temperature relative to the percentage change in the magnetic field. We plan to discuss these results in detail in a future study.

The current observations align with the Raghav hypothesis concerning coronal heating and solar wind generation. The Sun’s photospheric temperature is approximately 5,778 K (about 5,800 K). According to the Raghav effect, plasma particles can gain energy while in transit through successive reconnection events. Assuming the energy transfer efficiency is 100\%, each reconnection event effectively doubles the photospheric temperature of a particle, increasing it to approximately 11,600 K after the first event, 23,200 K after the second, and 46,400 K after the third. By the eighth reconnection cycle, the particle’s temperature can reach around 1.1 million K, consistent with observed coronal temperatures. Even with lower efficiency, only a modest increase in the number of reconnection events is needed to achieve coronal heating in such a turbulent environment.
Furthermore, if a plasma particle accumulates enough energy crossing through multiple reconnection events, it can exceed the solar escape velocity, contributing to solar wind formation. In summary, we opine that the magnetic reconnection process implied by the Raghav effect is the fundamental process contributing to the coronal heating and solar wind generation and their acceleration.

\begin{figure}
 \includegraphics[width=1\textwidth]{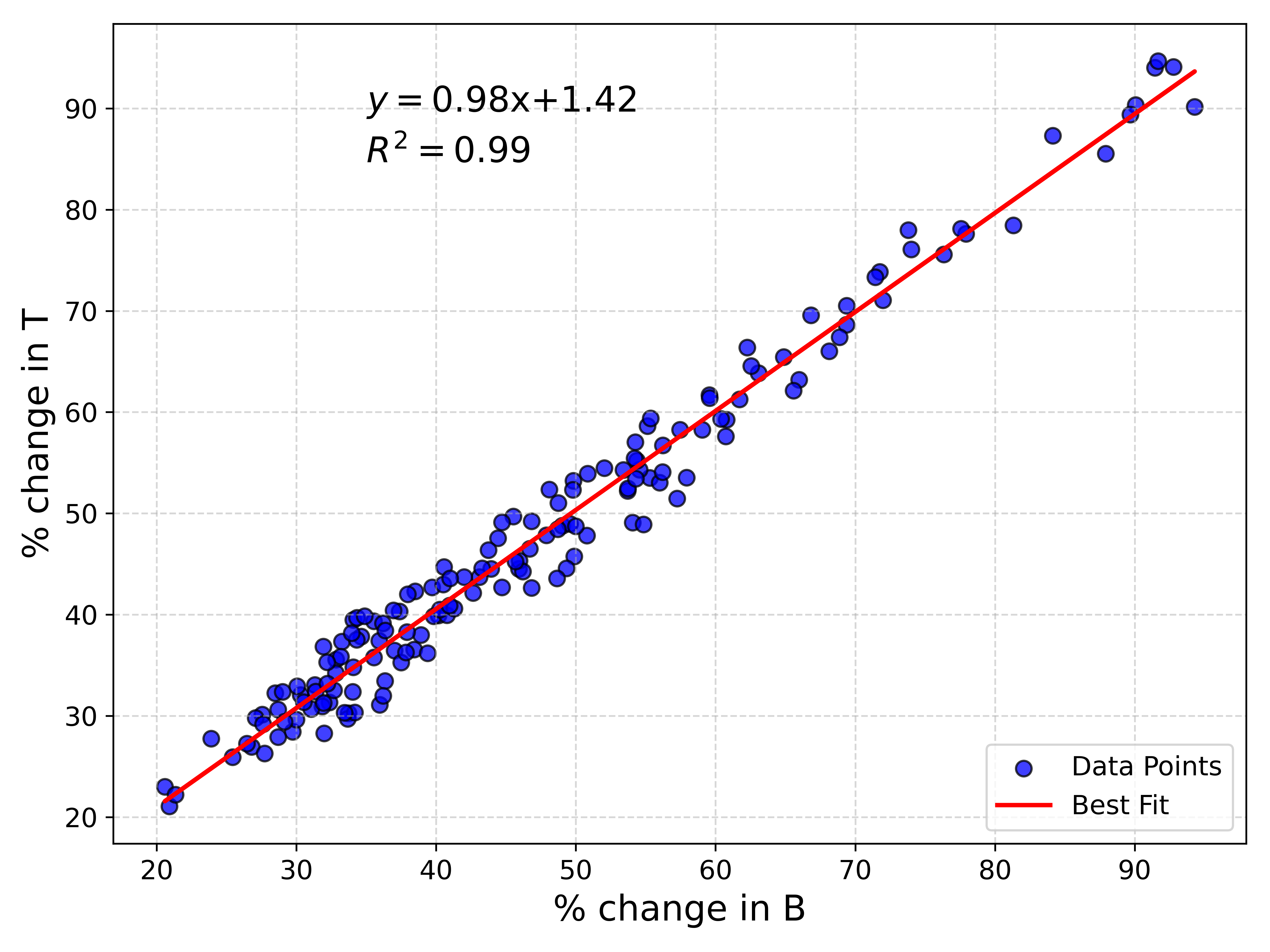}
  \caption{
The scatterplot displays the percentage change in temperature and magnetic field, highlighting a strong correlation. The data points (blue) are accompanied by a linear fit (red). } 
 \end{figure}

\section{Discussion and Implication}
Speiser (1965) demonstrated that particles can deviate from their expected trajectories in regions where magnetic fields change rapidly, resulting in unpredictable and chaotic motion \cite{speiser1965particle}.  Building on this foundational work, more recent studies by Drake et al. (2006) and Hoshino (2012) have shown that as particles enter the diffusion region—a critical zone in the reconnection process—they become unmagnetized \cite{drake2006electron,hoshino2012stochastic}. This loss of magnetic guidance allows particles to be directly influenced by the electric fields generated during reconnection, leading to significant acceleration.
Recently, EUV solar observations have revealed that small-scale magnetic reconnection at the base of the solar corona drives omnipresent jetting activity, which are responsible for heating and accelerating the solar wind \cite{raouafi2023magnetic}.
The study further suggests that these jets, occurring regardless of the solar cycle phase, produce intermittent outflows of hot plasma and Alfvén waves. Moreover, Wang et al. (2023) provides direct evidence of turbulent magnetic reconnection in the solar wind. Their findings show that this reconnection occurs within the solar wind's turbulent environment, influencing its dynamics \cite{wang2023direct}. Both these studies are linking small-scale reconnection to broader solar phenomena like coronal heating and solar wind acceleration.

In the past, theories based on acoustic and magnetohydrodynamic (MHD) waves, as well as micro/nano-flares, have been proposed to explain the extreme heating of the solar corona and the origins of the supersonic solar wind. Acoustic waves are believed to transfer energy upwards, but this energy is often insufficient to account for the high temperatures observed in the corona \cite{narain1990chromospheric}. MHD wave theories, particularly those involving Alfvén waves, suggest energy transport along magnetic field lines, yet the mechanisms by which these waves dissipate energy remain unclear \cite{cranmer2007self,cranmer2019properties}. Micro/nano-flares, which involve small-scale magnetic reconnection, are considered potential contributors, but direct evidence of their cumulative impact is limited \cite{parker1988nanoflares,klimchuk2006solving}. Recent studies suggest that small-scale magnetic reconnection and jetting activity may play a more significant role, highlighting the need for further research to fully understand these solar phenomena \cite{raouafi2023magnetic,wang2023direct}. Consequently, there is a growing consensus in the scientific community that magnetic reconnection is a fundamental process for plasma heating and energization. However, the exact mechanism of energy transfer to particles during magnetic reconnection remains ambiguous. While some researchers suggest that the reconnecting electric field contributes to energization, a direct relationship for this process is still elusive.

Raghav (2025) proposed a hypothesis that explores the  charged particles' magnetic moments-fields  coupling to address the missing link in energy transfer during magnetic reconnection \cite{raghav2025solarenigmadecodingmystery}. At the reconnection X-point, the magnetic field undergoes rapid reorientation over very short spatial and temporal scales. Particles, which previously gyrated around the field lines, retain a `memory' of their conserved magnetic moment just before reconnection. In this critical region, the field's variation is too rapid relative to the particle's gyro-motion. According to the Raghav hypothesis, this shift in magnetic field configuration prompts an adjustment in the magnetic moment alignment, resulting in a change in kinetic energy. This sudden energy increase alters the particles' gyrating motion, breaking the magnetic-invariant condition and causing a change in the magnetic moment. Consequently, particles experience chaotic motion, detaching from their original field lines. Some particles are expelled as exhaust, while others remain within the region, carrying increased energy and a modified magnetic moment.

The hypothesis posits that the relative change in the magnetic field is proportional to the relative change in temperature, as detailed in Equation 5. The empirical relationship, based on in-situ measurements and illustrated in Figure 2, demonstrates that changes in the magnetic field lead to the energization of plasma particles. This linear relationship provides crucial evidence supporting the hypothesis that variations in the magnetic field are directly linked to changes in particle energy. This dynamic interplay underscores the fundamental energy transfer processes occurring during magnetic reconnection and emphasizes the potential of the Raghav effect to enhance our understanding of plasma behavior in magnetic environments significantly.

However, it is important to recognize other contributing factors, such as distinct plasma waves, whether MHD or kinetic in nature, and turbulence dissipation, which also play roles in solar wind acceleration \cite{rivera2024situ,raouafi2023magnetic,wang2023direct}. While these effects are involved, their contribution is likely less significant in our selected events compared to the coupling of particle magnetic moments and field. Nonetheless, there are instances where the temperature increase is less pronounced than the decrease in the magnetic field, suggesting that waves or turbulence might be the dominant energy transfer processes in those cases. Currently, we lack a clear understanding of the criteria that determine which mode becomes dominant in energy exchange scenarios. This gap in knowledge highlights the complexity of the interactions within the solar wind and the need for further investigation. The dominance of either plasma waves or turbulence could depend on a variety of factors, such as the local plasma conditions, the scale of the magnetic field fluctuations, or the specific characteristics of the solar wind stream. Additionally, the interplay between different modes might be influenced by the initial conditions of the solar wind, such as its density, velocity, and temperature. Understanding these dynamics requires comprehensive observational data and advanced modeling efforts to unravel the intricate processes at play. Future research should aim to identify the conditions under which each mode prevails, potentially leading to a more nuanced understanding of solar wind acceleration and its variability.


Compelling observational evidence from in situ measurements of space plasma at 1 AU strongly supports the existence of the Raghav effect. The presence of plasma and magnetic fields across diverse environments—ranging from planetary magnetospheres and stellar surroundings, including the heliosphere, to the interstellar medium, galaxies, and the broader universe—suggests that the Raghav effect is a universal mechanism for particle energization and heating. The interaction between turbulent magnetic fields and charged particles may be crucial for driving phenomena such as cosmic ray acceleration, plasma heating, and explosive events like flares or gamma-ray bursts. Recognizing the Raghav effect not only enriches our comprehension of these processes but also provides a cohesive framework for exploring the intricate interplay of magnetic fields and particles in various cosmic environments.




\bmhead{Acknowledgements}
We acknowledge the Department of Physics, University of Mumbai for providing a facility for this work.



\begin{itemize}
\item Funding: NIL

\item Data availability: The data used in this analysis were obtained from the Wind spacecraft. They are publicly accessible at: (1) NASA’s Goddard Space Flight Center (GSFC) \url{https://wind.nasa.gov/data.php}, and (2) Coordinated Data Analysis Web (CDAWeb) \url{https://cdaweb.gsfc.nasa.gov/pub/data/wind/}.
\item Author contribution: AR proposed the scientific concept and determined the analysis algorithm. AK analyzed the Wind spacecraft data and created the figures for this article. SK assisted with data analysis. KG, MK, KK, and OD participated in discussions. AR prepared the initial draft of the article, which was reviewed and revised by all authors.

\end{itemize}







\begin{appendices}

\section{Quantification method}\label{secA1}

We identified events likely involving transits through potential reconnection sites, with a prototype event illustrated in Figure 1 using color shading. The exact start and end times for these regions are detailed in the supplementary event list. Selection was based on manual classification criteria: (1) a nearly constant IMF followed by a sharp decrease and subsequent recovery to the background level, and (2) a concurrent increase in plasma temperature and plasma beta. We determined the minimum and maximum values of the IMF strength and temperature data within these time frames. The percentage changes were calculated using the following formulas:

For the magnetic field:
\begin{equation}
    \%~change ~in~ B = \frac{B_{max}- B_{min}}{B_{max}} \times 100
\end{equation}

For the temperature:
\begin{equation}
    \%~change ~in~ T = \frac{T_{max}- T_{min}}{T_{min}} \times 100
\end{equation}

The same algorithm and mathematical approach were applied uniformly across all analyzed events. This consistency in methodology ensured that our findings related to the Raghav effect are robust and comparable across various scenarios.



\end{appendices}

\begin{landscape}
\begin{center}
\scriptsize
\begin{longtable}{cccccccccc}
\caption{The event list is used in the analysis.}\label{tab:spread_latex} \\
\toprule
\textbf{Sr. no.} &\textbf{dd-mm-yyyy} &\textbf{Start time} &\textbf{End time} &\textbf{B max} &\textbf{B min} &\textbf{T max} &\textbf{T min} &\textbf{\% change in B} &\textbf{\% change in T} \\
\midrule
\endfirsthead
\multicolumn{10}{c}%
{{\bfseries \tablename\ \thetable{} -- continued from previous page}} \\
\toprule
\textbf{Sr. no.}&\textbf{dd-mm-yyyy} & \textbf{Start Time} & \textbf{End Time} & \textbf{B max} & \textbf{B min} & \textbf{T max} & \textbf{T min} & \textbf{\% Change in B} & \textbf{\% Change in T} \\
\midrule
\endhead
\midrule \multicolumn{10}{c}{{Continued on next page}} \\
\endfoot
\bottomrule
\endlastfoot
1 &01-06-2008 &19:08:00 &19:11:00 &4.19 &2.1 &261539 &179456 &49.88 &45.74 \\
2 &01-06-2008 &15:22:00 &15:27:00 &4.16 &2.62 &282416 &206980 &37.02 &36.45 \\
3 &02-06-2008 &23:10:30 &23:13:00 &3.82 &1.5 &164864 &104597 &60.73 &57.62 \\
4 &04-06-2008 &03:20:00 &03:25:00 &3.4 &1.52 &159170 &103692 &55.29 &53.50 \\
5 &04-06-2008 &16:26:00 &16:28:00 &2.59 &1.32 &88022 &59166 &49.03 &48.77 \\
6 &04-06-2008 &16:28:42 &16:29:42 &2.59 &1.82 &79319 &61756 &29.73 &28.44 \\
7 &05-06-2008 &15:24:30 &15:30:00 &2.9 &0.64 &66118 &37225 &77.93 &77.62 \\
8 &08-06-2008 &00:19:00 &00:22:00 &4.62 &3.22 &182207 &137962 &30.30 &32.07 \\
9 &16-06-2008 &11:18:00 &11:22:00 &5.65 &3.14 &364522 &247052 &44.42 &47.55 \\
10 &20-06-2008 &18:56:00 &18:58:00 &4.5 &1.82 &226435 &140048 &59.56 &61.68 \\
11 &26-06-2008 &01:10:00 &01:17:00 &8.74 &4.9 &404473 &279892 &43.94 &44.51 \\
12 &01-07-2009 &14:16:00 &14:18:00 &5.1 &2.17 &99124 &62630 &57.45 &58.27 \\
13 &07-07-2009 &03:47:30 &03:49:39 &3.55 &2.02 &73091 &50860 &43.10 &43.71 \\
14 &09-07-2009 &23:15:00 &23:20:00 &3.51 &0.3 &44544 &22957 &91.45 &94.03 \\
15 &12-07-2009 &21:55:00 &21:57:00 &4.5 &2.28 &315563 &218287 &49.33 &44.56 \\
16 &12-07-2009 &22:10:00 &22:20:00 &4.96 &1.39 &313325 &183159 &71.98 &71.07 \\
17 &13-07-2009 &21:50:00 &21:51:30 &3.77 &2.43 &420418 &301717 &35.54 &39.34 \\
18 &03-08-2008 &17:42:00 &17:51:00 &5.19 &3.63 &80596.54 &60636.66 &30.06 &32.92 \\
19 &04-08-2008 &02:16:00 &02:21:00 &2.81 &1.41 &57314.96 &37406.06 &49.82 &53.22 \\
20 &04-08-2008 &11:00:00 &11:10:00 &3.54 &2.18 &61034.95 &44696.33 &38.42 &36.55 \\
21 &05-08-2008 &11:08:00 &11:11:00 &3.47 &2.54 &50853.98 &40064.97 &26.80 &26.93 \\
22 &05-08-2008 &16:18:00 &16:26:00 &3.92 &2.84 &66019.6 &50735.46 &27.55 &30.13 \\
23 &06-08-2008 &04:23:00 &04:27:00 &4.08 &1.3 &48671.93 &29316.2 &68.14 &66.02 \\
24 &06-08-2008 &07:55:00 &08:05:00 &4.39 &2.93 &82261.72 &59895.67 &33.26 &37.34 \\
25 &10-08-2008 &01:22:00 &01:27:00 &8.56 &5.79 &697443.81 &531055.7 &32.36 &31.33 \\
26 &11-08-2008 &00:17:00 &00:27:00 &4.07 &2.66 &335238.83 &243249.07 &34.64 &37.82 \\
27 &14-08-2008 &05:32:00 &05:42:00 &5.36 &3.42 &151467.17 &108867.96 &36.19 &39.13 \\
28 &16-08-2008 &18:06:00 &18:15:00 &5.09 &3.24 &63174.89 &47338.45 &36.35 &33.45 \\
29 &21-08-2008 &10:39:00 &10:45:00 &3.09 &2.21 &94535.87 &71492.18 &28.48 &32.23 \\
30 &21-08-2008 &14:49:00 &14:59:00 &3.22 &0.32 &98477.14 &51740.29 &90.06 &90.33 \\
31 &28-08-2008 &03:36:00 &03:44:00 &4.17 &2.96 &72793.17 &54988.74 &29.02 &32.38 \\
32 &01-09-2008 &11:25:00 &11:31:00 &2.34 &1.4 &46707 &33381 &40.17 &39.92 \\
33 &01-09-2008 &19:40:00 &19:45:00 &4 &1.13 &63472 &36506 &71.75 &73.87 \\
34 &06-09-2008 &10:33:36 &11:02:24 &6.13 &4.47 &219434 &169070 &27.08 &29.79 \\
35 &07-09-2008 &00:17:00 &00:21:00 &5.38 &3.12 &298743 &207867 &42.01 &43.72 \\
36 &07-09-2008 &19:44:00 &19:49:00 &4.52 &2.72 &240081 &171672 &39.82 &39.85 \\
37 &08-09-2008 &09:14:00 &09:17:00 &2.97 &2.15 &196208 &151885 &27.61 &29.18 \\
38 &10-09-2008 &03:25:00 &03:30:00 &3.38 &2.27 &135962 &100294 &32.84 &35.56 \\
39 &10-09-2008 &06:49:00 &06:50:00 &2.83 &2.11 &116379 &92419 &25.44 &25.92 \\
40 &10-09-2008 &09:09:00 &09:15:00 &3.78 &2.78 &100651 &79089 &26.46 &27.26 \\
41 &10-09-2008 &16:30:00 &16:35:00 &2.23 &1.02 &132162 &84161 &54.26 &57.04 \\
42 &12-09-2008 &03:13:00 &03:10:00 &2.51 &1.79 &90876 &69570 &28.69 &30.63 \\
43 &13-09-2008 &08:34:00 &08:38:00 &3.5 &1.34 &32095 &19903 &61.71 &61.26 \\
44 &15-09-2008 &07:04:00 &07:09:00 &12.02 &7.93 &444524 &335808 &34.03 &32.37 \\
45 &15-09-2008 &18:47:00 &18:53:00 &4.95 &2.25 &241840 &156723 &54.55 &54.31 \\
46 &16-09-2008 &01:25:00 &01:27:00 &3.3 &2.31 &187000 &144267 &30.00 &29.62 \\
47 &16-09-2008 &07:49:00 &07:55:00 &3.16 &1.93 &192350 &139388 &38.92 &38.00 \\
48 &23-09-2008 &10:47:00 &10:52:00 &4.37 &2.8 &60409 &43960 &35.93 &37.42 \\
49 &25-09-2008 &02:11:00 &02:16:00 &3.5 &1.54 &47761 &31209 &56.00 &53.04 \\
50 &25-09-2008 &15:30:00 &15:32:00 &5.3 &3.56 &96042 &71549 &32.83 &34.23 \\
51 &29-09-2008 &10:03:00 &10:08:00 &4.34 &2.9 &103675 &76321 &33.18 &35.84 \\
52 &01-10-2008 &07:54:00 &07:56:00 &6.45 &4.43 &259365 &194956 &31.32 &33.04 \\
53 &02-10-2008 &04:25:00 &04:27:00 &7.17 &5.08 &426296 &329357 &29.15 &29.43 \\
54 &02-10-2008 &08:11:30 &08:12:30 &7.43 &4.4 &461302 &329591 &40.78 &39.96 \\
55 &02-10-2008 &11:03:00 &11:04:00 &5.07 &4.01 &341213 &281886 &20.91 &21.05 \\
56 &02-10-2008 &20:53:00 &20:55:00 &3.75 &2.71 &284745 &225483 &27.73 &26.28 \\
57 &03-10-2008 &00:13:00 &00:17:00 &4.47 &1.57 &345201 &208639 &64.88 &65.45 \\
58 &03-10-2008 &00:40:00 &00:45:00 &4.8 &1.94 &352584 &218449 &59.58 &61.40 \\
59 &03-10-2008 &07:15:00 &07:17:00 &4.42 &3.51 &277102 &225309 &20.59 &22.99 \\
60 &04-10-2008 &03:01:00 &03:03:00 &4.45 &3.5 &304020 &248765 &21.35 &22.21 \\
61 &04-10-2008 &15:30:00 &15:32:00 &3.6 &2.47 &209508 &158235 &31.39 &32.40 \\
62 &05-10-2008 &10:48:00 &10:52:00 &3.55 &2.34 &178438 &127944 &34.08 &39.47 \\
63 &07-10-2008 &21:50:00 &21:52:00 &3.85 &0.22 &106606 &56063 &94.29 &90.15 \\
64 &07-10-2008 &22:18:00 &22:22:00 &3.04 &0.79 &88114 &50042 &74.01 &76.08 \\
65 &07-10-2008 &22:24:00 &22:28:00 &3.05 &0.57 &93330 &52297 &81.31 &78.46 \\
66 &08-10-2008 &04:10:00 &04:16:00 &3.88 &1.32 &69939 &42856 &65.98 &63.19 \\
67 &09-10-2008 &10:30:00 &10:34:00 &2.87 &1.31 &48979 &31558 &54.36 &55.20 \\
68 &12-10-2008 &17:06:00 &17:08:00 &3.6 &2.17 &257226 &180268 &39.72 &42.69 \\
69 &13-10-2008 &22:34:00 &22:42:00 &2.92 &1.99 &137045 &104669 &31.85 &30.93 \\
70 &17-10-2008 &23:22:00 &23:25:00 &2.04 &1.17 &25538 &17966 &42.65 &42.14 \\
71 &19-10-2008 &23:04:00 &23:12:00 &7 &3.97 &196805 &136144 &43.29 &44.56 \\
72 &20-10-2008 &20:04:00 &20:12:00 &6.34 &3.97 &183281 &130624 &37.38 &40.31 \\
73 &21-10-2008 &08:22:00 &08:26:00 &4.32 &2.57 &121457 &84943 &40.51 &42.99 \\
74 &29-10-2008 &01:30:00 &01:37:00 &8.48 &5.62 &484251 &371760 &33.73 &30.26 \\
75 &29-10-2008 &04:10:00 &04:16:00 &8.58 &3.93 &464897 &299074 &54.20 &55.45 \\
76 &02-11-2008 &06:14:45 &06:20:00 &4.75 &2.08 &173219 &112413 &56.21 &54.09 \\
77 &02-11-2008 &12:57:00 &12:58:00 &3.32 &1.49 &164627 &103775 &55.12 &58.64 \\
78 &06-11-2008 &17:37:00 &17:39:00 &4.13 &2.54 &24054 &16902 &38.50 &42.31 \\
79 &08-11-2008 &00:51:00 &00:55:00 &6.21 &3.71 &331999 &236312 &40.26 &40.49 \\
80 &08-11-2008 &19:25:51 &19:30:00 &4.77 &1.46 &301294 &176694 &69.39 &70.52 \\
81 &11-11-2008 &19:47:18 &19:50:00 &2.46 &1.33 &76772 &53125 &45.93 &44.51 \\
82 &12-11-2008 &20:50:00 &21:00:00 &2.83 &1.21 &46150 &30467 &57.24 &51.48 \\
83 &17-11-2008 &15:35:00 &15:40:00 &3.84 &2.16 &124840 &85287 &43.75 &46.38 \\
84 &22-11-2008 &07:47:00 &07:50:30 &1.99 &1.32 &27739 &21387 &33.67 &29.70 \\
85 &22-11-2008 &15:16:00 &15:22:00 &2.64 &1.68 &25633 &18515 &36.36 &38.44 \\
86 &26-11-2008 &15:17:00 &15:19:00 &5.14 &2.38 &298825 &196311 &53.70 &52.22 \\
87 &26-11-2008 &19:55:00 &19:57:30 &4.32 &3.08 &319218 &249588 &28.70 &27.90 \\
88 &27-11-2008 &17:55:12 &17:55:30 &2.93 &2.02 &215195 &164679 &31.06 &30.68 \\
89 &28-11-2008 &06:38:00 &06:43:00 &3.03 &1.99 &101516 &73822 &34.32 &37.51 \\
90 &28-11-2008 &07:03:00 &07:05:00 &2.7 &1.24 &117813 &79028 &54.07 &49.08 \\
91 &01-12-2008 &12:34:00 &12:40:00 &3.1 &0.32 &41976 &22161 &89.68 &89.41 \\
92 &01-12-2008 &11:30:00 &11:40:00 &3.32 &0.24 &40047 &20632 &92.77 &94.11 \\
93 &01-12-2008 &16:16:00 &16:26:00 &3.05 &1.05 &32641 &20134 &65.57 &62.12 \\
94 &06-12-2008 &02:07:00 &02:09:39 &4.36 &2.81 &204491 &150627 &35.55 &35.76 \\
95 &06-12-2008 &02:52:00 &02:53:15 &4.03 &2.5 &214880 &151324 &37.97 &42.00 \\
96 &06-12-2008 &19:59:30 &20:02:00 &3.78 &2.01 &179409 &120238 &46.83 &49.21 \\
97 &09-12-2008 &01:51:00 &01:55:00 &2.49 &1.48 &77614 &53634 &40.56 &44.71 \\
98 &10-12-2008 &10:02:15 &10:02:45 &4.37 &1.45 &55840 &32933 &66.82 &69.56 \\
99 &10-12-2008 &11:50:00 &11:58:27 &4.14 &0.5 &69770 &37603 &87.92 &85.54 \\
100 &16-12-2008 &00:10:00 &00:25:00 &3.53 &0.56 &58123 &31028 &84.14 &87.32 \\
101 &16-12-2008 &21:56:30 &21:59:30 &5.83 &3.62 &73760 &53346 &37.91 &38.27 \\
102 &19-12-2008 &12:05:00 &12:10:00 &4.34 &2.85 &34962 &25030 &34.33 &39.68 \\
103 &24-12-2008 &08:10:00 &08:16:00 &4.41 &2.17 &203784 &137876 &50.79 &47.80 \\
104 &24-12-2008 &11:46:00 &11:52:00 &4.81 &2.62 &197874 &132183 &45.53 &49.70 \\
105 &24-12-2008 &12:01:00 &12:02:00 &5.07 &3.45 &179784 &136961 &31.95 &31.27 \\
106 &24-12-2008 &12:52:00 &12:53:00 &4.44 &3.02 &197385 &153899 &31.98 &28.26 \\
107 &25-12-2008 &08:10:00 &08:16:00 &3.31 &1.76 &141531 &99227 &46.83 &42.63 \\
108 &25-12-2008 &18:26:00 &18:34:00 &3.3 &2 &134163 &98513 &39.39 &36.19 \\
109 &27-12-2008 &01:52:45 &01:55:00 &2.38 &1.62 &66086 &48295 &31.93 &36.84 \\
110 &27-12-2008 &12:14:00 &12:15:30 &2.76 &2.1 &56251 &44035 &23.91 &27.74 \\
111 &27-12-2008 &21:39:00 &21:45:00 &2.8 &1.75 &58850 &43504 &37.50 &35.27 \\
112 &02-01-2009 &17:00:00 &17:20:00 &5.84 &1.53 &261603 &146974 &73.80 &77.99 \\
113 &08-01-2009 &01:06:00 &01:10:00 &3.33 &2.07 &38886 &28539 &37.84 &36.26 \\
114 &15-01-2009 &10:10:00 &10:15:00 &4.86 &1.09 &110467 &62018 &77.57 &78.12 \\
115 &29-01-2009 &06:50:00 &07:02:00 &5.44 &2.94 &157133 &108120 &45.96 &45.33 \\
116 &30-01-2009 &11:02:15 &11:03:45 &3.64 &2.45 &91708 &69201 &32.69 &32.52 \\
117 &03-02-2009 &05:35:00 &05:39:00 &1.88 &0.77 &38977 &24630 &59.04 &58.25 \\
118 &03-02-2009 &05:49:30 &05:50:15 &1.96 &1.29 &30504 &23405 &34.18 &30.33 \\
119 &03-03-2009 &19:16:45 &19:17:51 &6.84 &4.38 &129904 &99084 &35.96 &31.10 \\
120 &20-03-2009 &18:00:00 &18:02:00 &7.37 &5.12 &121021 &92131 &30.53 &31.36 \\
121 &20-03-2009 &01:25:00 &01:28:00 &3.89 &1.8 &47777 &31338 &53.73 &52.46 \\
122 &22-03-2009 &21:04:00 &21:04:45 &3.53 &1.78 &132907 &89229 &49.58 &48.95 \\
123 &23-03-2009 &14:00:00 &14:05:00 &2.88 &0.24 &106626 &54771 &91.67 &94.68 \\
124 &23-03-2009 &14:13:00 &14:18:30 &3.08 &0.88 &107913 &62257 &71.43 &73.33 \\
125 &23-03-2009 &10:09:00 &10:14:00 &2.69 &1.29 &102105 &66094 &52.04 &54.48 \\
126 &27-03-2009 &12:43:00 &12:45:00 &4.69 &2.5 &120049 &81945 &46.70 &46.50 \\
127 &27-03-2009 &12:48:00 &12:49:00 &4.77 &2.8 &113429 &80671 &41.30 &40.61 \\
128 &28-03-2009 &16:36:00 &16:41:00 &4.44 &1.36 &138373 &82053 &69.37 &68.64 \\
129 &31-03-2009 &13:40:00 &13:50:00 &3.59 &1.84 &95989 &63562 &48.75 &51.02 \\
130 &01-04-2009 &10:11:00 &10:16:30 &3.16 &1.7 &59265 &41086 &46.20 &44.25 \\
131 &01-04-2009 &13:57:00 &14:05:00 &2.64 &1.23 &57863 &37505 &53.41 &54.28 \\
132 &03-04-2009 &12:43:00 &12:52:00 &2.32 &0.91 &23446 &14723 &60.78 &59.25 \\
133 &03-04-2009 &17:33:00 &17:40:00 &2.93 &1.62 &59074 &41401 &44.71 &42.69 \\
134 &05-04-2009 &18:11:00 &18:17:00 &5.62 &3.81 &118674 &89108 &32.21 &33.18 \\
135 &10-04-2009 &13:20:00 &13:23:00 &4.26 &2.22 &194759 &131734 &47.89 &47.84 \\
136 &10-04-2009 &20:03:00 &20:03:51 &4.07 &2.09 &221797 &154462 &48.65 &43.59 \\
137 &12-04-2009 &01:37:00 &01:49:00 &3.08 &1.58 &171927 &115815 &48.70 &48.45 \\
138 &12-04-2009 &06:56:00 &06:59:00 &3.31 &1.83 &152259 &102104 &44.71 &49.12 \\
139 &12-04-2009 &15:14:30 &15:15:51 &3.21 &2.09 &115632 &82682 &34.89 &39.85 \\
140 &19-04-2009 &05:55:00 &06:02:00 &3.79 &1.43 &150240 &90299 &62.27 &66.38 \\
141 &19-04-2009 &21:51:24 &21:55:00 &4.09 &1.51 &140424 &85709 &63.08 &63.84 \\
142 &22-04-2009 &02:50:00 &03:00:00 &2.67 &1.76 &97823 &72562 &34.08 &34.81 \\
143 &23-04-2009 &03:34:39 &03:37:00 &2.39 &1.41 &72997 &50841 &41.00 &43.58 \\
144 &23-04-2009 &11:43:30 &11:46:30 &2.1 &0.96 &55853 &36401 &54.29 &53.44 \\
145 &27-04-2009 &17:56:00 &17:58:00 &3.01 &1.92 &63561 &48162 &36.21 &31.97 \\
146 &27-04-2009 &19:48:00 &19:58:00 &2.93 &1.44 &64808 &42103 &50.85 &53.93 \\
147 &04-05-2009 &03:01:00 &03:01:45 &3.76 &2.22 &96147 &68245 &40.96 &40.89 \\
148 &04-05-2009 &23:50:00 &23:52:00 &2.22 &1.4 &45179 &32172 &36.94 &40.43 \\
149 &04-05-2009 &23:53:00 &23:57:54 &2.21 &0.93 &52137 &33960 &57.92 &53.52 \\
150 &07-05-2009 &05:24:00 &05:30:00 &4.49 &2.33 &145246 &95325 &48.11 &52.37 \\
151 &07-05-2009 &14:47:00 &14:49:00 &4.03 &1.82 &113208 &76028 &54.84 &48.90 \\
152 &07-05-2009 &16:09:45 &16:17:00 &5.15 &2.04 &111988 &70287 &60.39 &59.33 \\
153 &08-05-2009 &14:08:00 &14:10:30 &4.56 &2.29 &188774 &123913 &49.78 &52.34 \\
154 &08-05-2009 &14:49:51 &14:52:00 &4.62 &2.51 &199275 &137204 &45.67 &45.24 \\
155 &09-05-2009 &11:59:00 &12:02:00 &3.71 &2.45 &159053 &115126 &33.96 &38.16 \\
156 &09-05-2009 &19:40:54 &19:48:00 &2.33 &1.04 &117496 &73722 &55.36 &59.38 \\
157 &11-05-2009 &12:27:00 &12:28:30 &3.14 &1.57 &127259 &85566 &50.00 &48.73 \\
158 &12-05-2009 &13:49:00 &13:53:00 &2.95 &2 &61515 &45470 &32.20 &35.29 \\
159 &12-05-2009 &20:17:00 &20:19:00 &2.87 &1.91 &52092 &39976 &33.45 &30.31 \\
160 &15-05-2009 &22:43:30 &22:50:00 &3.47 &1.08 &83923 &50134 &68.88 &67.40 \\
161 &15-05-2009 &23:03:00 &23:07:00 &3.93 &0.93 &80226 &45689 &76.34 &75.59 \\
162 &16-05-2009 &20:40:00 &20:47:00 &3.31 &1.24 &109125 &66317 &62.54 &64.55 \\
163 &16-05-2009 &20:54:00 &21:03:00 &3.7 &1.62 &106623 &68032 &56.22 &56.72 \\
\bottomrule
\end{longtable}
\end{center}
\end{landscape}

\bibliography{Corona.bib}


\begin{thebibliography}{27}
\ifx \bisbn   \undefined \def \bisbn  #1{ISBN #1}\fi
\ifx \binits  \undefined \def \binits#1{#1}\fi
\ifx \bauthor  \undefined \def \bauthor#1{#1}\fi
\ifx \batitle  \undefined \def \batitle#1{#1}\fi
\ifx \bjtitle  \undefined \def \bjtitle#1{#1}\fi
\ifx \bvolume  \undefined \def \bvolume#1{\textbf{#1}}\fi
\ifx \byear  \undefined \def \byear#1{#1}\fi
\ifx \bissue  \undefined \def \bissue#1{#1}\fi
\ifx \bfpage  \undefined \def \bfpage#1{#1}\fi
\ifx \blpage  \undefined \def \blpage #1{#1}\fi
\ifx \burl  \undefined \def \burl#1{\textsf{#1}}\fi
\ifx \doiurl  \undefined \def \doiurl#1{\url{https://doi.org/#1}}\fi
\ifx \betal  \undefined \def \betal{\textit{et al.}}\fi
\ifx \binstitute  \undefined \def \binstitute#1{#1}\fi
\ifx \binstitutionaled  \undefined \def \binstitutionaled#1{#1}\fi
\ifx \bctitle  \undefined \def \bctitle#1{#1}\fi
\ifx \beditor  \undefined \def \beditor#1{#1}\fi
\ifx \bpublisher  \undefined \def \bpublisher#1{#1}\fi
\ifx \bbtitle  \undefined \def \bbtitle#1{#1}\fi
\ifx \bedition  \undefined \def \bedition#1{#1}\fi
\ifx \bseriesno  \undefined \def \bseriesno#1{#1}\fi
\ifx \blocation  \undefined \def \blocation#1{#1}\fi
\ifx \bsertitle  \undefined \def \bsertitle#1{#1}\fi
\ifx \bsnm \undefined \def \bsnm#1{#1}\fi
\ifx \bsuffix \undefined \def \bsuffix#1{#1}\fi
\ifx \bparticle \undefined \def \bparticle#1{#1}\fi
\ifx \barticle \undefined \def \barticle#1{#1}\fi
\bibcommenthead
\ifx \bconfdate \undefined \def \bconfdate #1{#1}\fi
\ifx \botherref \undefined \def \botherref #1{#1}\fi
\ifx \url \undefined \def \url#1{\textsf{#1}}\fi
\ifx \bchapter \undefined \def \bchapter#1{#1}\fi
\ifx \bbook \undefined \def \bbook#1{#1}\fi
\ifx \bcomment \undefined \def \bcomment#1{#1}\fi
\ifx \oauthor \undefined \def \oauthor#1{#1}\fi
\ifx \citeauthoryear \undefined \def \citeauthoryear#1{#1}\fi
\ifx \endbibitem  \undefined \def \endbibitem {}\fi
\ifx \bconflocation  \undefined \def \bconflocation#1{#1}\fi
\ifx \arxivurl  \undefined \def \arxivurl#1{\textsf{#1}}\fi
\csname PreBibitemsHook\endcsname

\bibitem[\protect\citeauthoryear{Klimchuk}{2006}]{klimchuk2006solving}
\begin{barticle}
\bauthor{\bsnm{Klimchuk}, \binits{J.A.}}:
\batitle{On solving the coronal heating problem}.
\bjtitle{Solar Physics}
\bvolume{234}(\bissue{1}),
\bfpage{41}--\blpage{77}
(\byear{2006})
\end{barticle}
\endbibitem

\bibitem[\protect\citeauthoryear{Aschwanden}{2006}]{aschwanden2006physics}
\begin{bbook}
\bauthor{\bsnm{Aschwanden}, \binits{M.}}:
\bbtitle{Physics of the Solar Corona: an Introduction with Problems and Solutions}.
\bpublisher{Springer}, \blocation{???}
(\byear{2006})
\end{bbook}
\endbibitem

\bibitem[\protect\citeauthoryear{McComas et~al.}{2007}]{mccomas2007understanding}
\begin{botherref}
\oauthor{\bsnm{McComas}, \binits{D.}},
\oauthor{\bsnm{Velli}, \binits{M.}},
\oauthor{\bsnm{Lewis}, \binits{W.}},
\oauthor{\bsnm{Acton}, \binits{L.}},
\oauthor{\bsnm{Balat-Pichelin}, \binits{M.}},
\oauthor{\bsnm{Bothmer}, \binits{V.}},
\oauthor{\bsnm{Dirling~Jr}, \binits{R.}},
\oauthor{\bsnm{Feldman}, \binits{W.}},
\oauthor{\bsnm{Gloeckler}, \binits{G.}},
\oauthor{\bsnm{Habbal}, \binits{S.}}, et al.:
Understanding coronal heating and solar wind acceleration: Case for in situ near-sun measurements.
Reviews of Geophysics
\textbf{45}(1)
(2007)
\end{botherref}
\endbibitem

\bibitem[\protect\citeauthoryear{Cranmer et~al.}{2007}]{cranmer2007self}
\begin{barticle}
\bauthor{\bsnm{Cranmer}, \binits{S.R.}},
\bauthor{\bsnm{Van~Ballegooijen}, \binits{A.A.}},
\bauthor{\bsnm{Edgar}, \binits{R.J.}}:
\batitle{Self-consistent coronal heating and solar wind acceleration from anisotropic magnetohydrodynamic turbulence}.
\bjtitle{The Astrophysical Journal Supplement Series}
\bvolume{171}(\bissue{2}),
\bfpage{520}
(\byear{2007})
\end{barticle}
\endbibitem

\bibitem[\protect\citeauthoryear{Cranmer and Winebarger}{2019}]{cranmer2019properties}
\begin{barticle}
\bauthor{\bsnm{Cranmer}, \binits{S.R.}},
\bauthor{\bsnm{Winebarger}, \binits{A.R.}}:
\batitle{The properties of the solar corona and its connection to the solar wind}.
\bjtitle{Annual Review of Astronomy and Astrophysics}
\bvolume{57},
\bfpage{157}--\blpage{187}
(\byear{2019})
\end{barticle}
\endbibitem

\bibitem[\protect\citeauthoryear{Lin et~al.}{1984}]{lin1984solar}
\begin{barticle}
\bauthor{\bsnm{Lin}, \binits{R.}},
\bauthor{\bsnm{Schwartz}, \binits{R.}},
\bauthor{\bsnm{Kane}, \binits{S.R.}},
\bauthor{\bsnm{Pelling}, \binits{R.}},
\bauthor{\bsnm{Hurley}, \binits{K.}}:
\batitle{Solar hard x-ray microflares}.
\bjtitle{The Astrophysical Journal}
\bvolume{283},
\bfpage{421}--\blpage{425}
(\byear{1984})
\end{barticle}
\endbibitem

\bibitem[\protect\citeauthoryear{Parker}{1988}]{parker1988nanoflares}
\begin{barticle}
\bauthor{\bsnm{Parker}, \binits{E.}}:
\batitle{Nanoflares and the solar x-ray corona}.
\bjtitle{The Astrophysical Journal}
\bvolume{330},
\bfpage{474}--\blpage{479}
(\byear{1988})
\end{barticle}
\endbibitem

\bibitem[\protect\citeauthoryear{Shibata and Magara}{2011}]{shibata2011solar}
\begin{barticle}
\bauthor{\bsnm{Shibata}, \binits{K.}},
\bauthor{\bsnm{Magara}, \binits{T.}}:
\batitle{Solar flares: magnetohydrodynamic processes}.
\bjtitle{Living Reviews in Solar Physics}
\bvolume{8}(\bissue{1}),
\bfpage{6}
(\byear{2011})
\end{barticle}
\endbibitem

\bibitem[\protect\citeauthoryear{Spitzer}{2006}]{spitzer2006physics}
\begin{bbook}
\bauthor{\bsnm{Spitzer}, \binits{L.}}:
\bbtitle{Physics of Fully Ionized Gases}.
\bpublisher{Courier Corporation}, \blocation{???}
(\byear{2006})
\end{bbook}
\endbibitem

\bibitem[\protect\citeauthoryear{Osterbrock}{1961}]{osterbrock1961heating}
\begin{barticle}
\bauthor{\bsnm{Osterbrock}, \binits{D.E.}}:
\batitle{The heating of the solar chromosphere, plages, and corona by magnetohydrodynamic waves.}
\bjtitle{The Astrophysical Journal}
\bvolume{134},
\bfpage{347}
(\byear{1961})
\end{barticle}
\endbibitem

\bibitem[\protect\citeauthoryear{Heyvaerts and Priest}{1983}]{heyvaerts1983coronal}
\begin{barticle}
\bauthor{\bsnm{Heyvaerts}, \binits{J.}},
\bauthor{\bsnm{Priest}, \binits{E.}}:
\batitle{Coronal heating by phase-mixed shear alfv{\'e}n waves}.
\bjtitle{Astronomy and Astrophysics}
\bvolume{117},
\bfpage{220}--\blpage{234}
(\byear{1983})
\end{barticle}
\endbibitem

\bibitem[\protect\citeauthoryear{Davila}{1987}]{davila1987heating}
\begin{barticle}
\bauthor{\bsnm{Davila}, \binits{J.M.}}:
\batitle{Heating of the solar corona by the resonant absorption of alfv{\'e}n waves}.
\bjtitle{The Astrophysical Journal}
\bvolume{317},
\bfpage{514}--\blpage{521}
(\byear{1987})
\end{barticle}
\endbibitem

\bibitem[\protect\citeauthoryear{Sakurai and Granik}{1984}]{sakurai1984generation}
\begin{barticle}
\bauthor{\bsnm{Sakurai}, \binits{T.}},
\bauthor{\bsnm{Granik}, \binits{A.}}:
\batitle{Generation of coronal electric currents due to convective motions on the photosphere. ii-resonance and phase mixing of alfven waves}.
\bjtitle{The Astrophysical Journal}
\bvolume{277},
\bfpage{404}--\blpage{414}
(\byear{1984})
\end{barticle}
\endbibitem

\bibitem[\protect\citeauthoryear{Gary}{1993}]{gary1993theory}
\begin{bbook}
\bauthor{\bsnm{Gary}, \binits{S.P.}}:
\bbtitle{Theory of Space Plasma Microinstabilities}
vol. \bseriesno{7}.
\bpublisher{Cambridge university press}, \blocation{???}
(\byear{1993})
\end{bbook}
\endbibitem

\bibitem[\protect\citeauthoryear{Gro{\v{s}}elj et~al.}{2019}]{grovselj2019kinetic}
\begin{barticle}
\bauthor{\bsnm{Gro{\v{s}}elj}, \binits{D.}},
\bauthor{\bsnm{Chen}, \binits{C.H.}},
\bauthor{\bsnm{Mallet}, \binits{A.}},
\bauthor{\bsnm{Samtaney}, \binits{R.}},
\bauthor{\bsnm{Schneider}, \binits{K.}},
\bauthor{\bsnm{Jenko}, \binits{F.}}:
\batitle{Kinetic turbulence in astrophysical plasmas: waves and/or structures?}
\bjtitle{Physical Review X}
\bvolume{9}(\bissue{3}),
\bfpage{031037}
(\byear{2019})
\end{barticle}
\endbibitem

\bibitem[\protect\citeauthoryear{Marsch}{2006}]{marsch2006kinetic}
\begin{barticle}
\bauthor{\bsnm{Marsch}, \binits{E.}}:
\batitle{Kinetic physics of the solar corona and solar wind}.
\bjtitle{Living Reviews in Solar Physics}
\bvolume{3},
\bfpage{1}--\blpage{100}
(\byear{2006})
\end{barticle}
\endbibitem

\bibitem[\protect\citeauthoryear{Bruno and Carbone}{2013}]{bruno2013solar}
\begin{barticle}
\bauthor{\bsnm{Bruno}, \binits{R.}},
\bauthor{\bsnm{Carbone}, \binits{V.}}:
\batitle{The solar wind as a turbulence laboratory}.
\bjtitle{Living Reviews in Solar Physics}
\bvolume{10}(\bissue{1}),
\bfpage{2}
(\byear{2013})
\end{barticle}
\endbibitem

\bibitem[\protect\citeauthoryear{Raghav}{2025}]{raghav2025solarenigmadecodingmystery}
\begin{botherref}
\oauthor{\bsnm{Raghav}, \binits{A.N.}}:
The Solar Enigma Decoding the Mystery of the Hot Corona and Solar Wind Acceleration
(2025).
\url{https://arxiv.org/abs/2003.10326}
\end{botherref}
\endbibitem

\bibitem[\protect\citeauthoryear{Lepping et~al.}{1995}]{lepping1995wind}
\begin{barticle}
\bauthor{\bsnm{Lepping}, \binits{R.}},
\bauthor{\bsnm{Ac{\~u}na}, \binits{M.}},
\bauthor{\bsnm{Burlaga}, \binits{L.}},
\bauthor{\bsnm{Farrell}, \binits{W.}},
\bauthor{\bsnm{Slavin}, \binits{J.}},
\bauthor{\bsnm{Schatten}, \binits{K.}},
\bauthor{\bsnm{Mariani}, \binits{F.}},
\bauthor{\bsnm{Ness}, \binits{N.}},
\bauthor{\bsnm{Neubauer}, \binits{F.}},
\bauthor{\bsnm{Whang}, \binits{Y.}}, \betal:
\batitle{The wind magnetic field investigation}.
\bjtitle{Space Science Reviews}
\bvolume{71},
\bfpage{207}--\blpage{229}
(\byear{1995})
\end{barticle}
\endbibitem

\bibitem[\protect\citeauthoryear{Lin et~al.}{1995}]{lin1995three}
\begin{barticle}
\bauthor{\bsnm{Lin}, \binits{R.}},
\bauthor{\bsnm{Anderson}, \binits{K.}},
\bauthor{\bsnm{Ashford}, \binits{S.}},
\bauthor{\bsnm{Carlson}, \binits{C.}},
\bauthor{\bsnm{Curtis}, \binits{D.}},
\bauthor{\bsnm{Ergun}, \binits{R.}},
\bauthor{\bsnm{Larson}, \binits{D.}},
\bauthor{\bsnm{McFadden}, \binits{J.}},
\bauthor{\bsnm{McCarthy}, \binits{M.}},
\bauthor{\bsnm{Parks}, \binits{G.}}, \betal:
\batitle{A three-dimensional plasma and energetic particle investigation for the wind spacecraft}.
\bjtitle{Space Science Reviews}
\bvolume{71},
\bfpage{125}--\blpage{153}
(\byear{1995})
\end{barticle}
\endbibitem

\bibitem[\protect\citeauthoryear{Speiser}{1965}]{speiser1965particle}
\begin{barticle}
\bauthor{\bsnm{Speiser}, \binits{T.}}:
\batitle{Particle trajectories in model current sheets: 1. analytical solutions}.
\bjtitle{Journal of Geophysical Research}
\bvolume{70}(\bissue{17}),
\bfpage{4219}--\blpage{4226}
(\byear{1965})
\end{barticle}
\endbibitem

\bibitem[\protect\citeauthoryear{Drake et~al.}{2006}]{drake2006electron}
\begin{barticle}
\bauthor{\bsnm{Drake}, \binits{J.}},
\bauthor{\bsnm{Swisdak}, \binits{M.}},
\bauthor{\bsnm{Che}, \binits{H.}},
\bauthor{\bsnm{Shay}, \binits{M.}}:
\batitle{Electron acceleration from contracting magnetic islands during reconnection}.
\bjtitle{Nature}
\bvolume{443}(\bissue{7111}),
\bfpage{553}--\blpage{556}
(\byear{2006})
\end{barticle}
\endbibitem

\bibitem[\protect\citeauthoryear{Hoshino}{2012}]{hoshino2012stochastic}
\begin{barticle}
\bauthor{\bsnm{Hoshino}, \binits{M.}}:
\batitle{Stochastic particle acceleration in multiple magnetic islands during reconnection}.
\bjtitle{Physical Review Letters}
\bvolume{108}(\bissue{13}),
\bfpage{135003}
(\byear{2012})
\end{barticle}
\endbibitem

\bibitem[\protect\citeauthoryear{Raouafi et~al.}{2023}]{raouafi2023magnetic}
\begin{barticle}
\bauthor{\bsnm{Raouafi}, \binits{N.E.}},
\bauthor{\bsnm{Stenborg}, \binits{G.}},
\bauthor{\bsnm{Seaton}, \binits{D.B.}},
\bauthor{\bsnm{Wang}, \binits{H.}},
\bauthor{\bsnm{Wang}, \binits{J.}},
\bauthor{\bsnm{DeForest}, \binits{C.E.}},
\bauthor{\bsnm{Bale}, \binits{S.D.}},
\bauthor{\bsnm{Drake}, \binits{J.F.}},
\bauthor{\bsnm{Uritsky}, \binits{V.M.}},
\bauthor{\bsnm{Karpen}, \binits{J.T.}}, \betal:
\batitle{Magnetic reconnection as the driver of the solar wind}.
\bjtitle{The Astrophysical Journal}
\bvolume{945}(\bissue{1}),
\bfpage{28}
(\byear{2023})
\end{barticle}
\endbibitem

\bibitem[\protect\citeauthoryear{Wang et~al.}{2023}]{wang2023direct}
\begin{barticle}
\bauthor{\bsnm{Wang}, \binits{R.}},
\bauthor{\bsnm{Wang}, \binits{S.}},
\bauthor{\bsnm{Lu}, \binits{Q.}},
\bauthor{\bsnm{Li}, \binits{X.}},
\bauthor{\bsnm{Lu}, \binits{S.}},
\bauthor{\bsnm{Gonzalez}, \binits{W.}}:
\batitle{Direct observation of turbulent magnetic reconnection in the solar wind}.
\bjtitle{Nature Astronomy}
\bvolume{7}(\bissue{1}),
\bfpage{18}--\blpage{28}
(\byear{2023})
\end{barticle}
\endbibitem

\bibitem[\protect\citeauthoryear{Narain and Ulmschneider}{1990}]{narain1990chromospheric}
\begin{barticle}
\bauthor{\bsnm{Narain}, \binits{U.}},
\bauthor{\bsnm{Ulmschneider}, \binits{P.}}:
\batitle{Chromospheric and coronal heating mechanisms}.
\bjtitle{Space Science Reviews}
\bvolume{54}(\bissue{3-4}),
\bfpage{377}--\blpage{445}
(\byear{1990})
\end{barticle}
\endbibitem

\bibitem[\protect\citeauthoryear{Rivera et~al.}{2024}]{rivera2024situ}
\begin{barticle}
\bauthor{\bsnm{Rivera}, \binits{Y.J.}},
\bauthor{\bsnm{Badman}, \binits{S.T.}},
\bauthor{\bsnm{Stevens}, \binits{M.L.}},
\bauthor{\bsnm{Verniero}, \binits{J.L.}},
\bauthor{\bsnm{Stawarz}, \binits{J.E.}},
\bauthor{\bsnm{Shi}, \binits{C.}},
\bauthor{\bsnm{Raines}, \binits{J.M.}},
\bauthor{\bsnm{Paulson}, \binits{K.W.}},
\bauthor{\bsnm{Owen}, \binits{C.J.}},
\bauthor{\bsnm{Niembro}, \binits{T.}}, \betal:
\batitle{In situ observations of large-amplitude alfv{\'e}n waves heating and accelerating the solar wind}.
\bjtitle{Science}
\bvolume{385}(\bissue{6712}),
\bfpage{962}--\blpage{966}
(\byear{2024})
\end{barticle}
\endbibitem

\end{thebibliography}

\end{document}